\documentclass[final,5p]{elsarticle}

\usepackage{graphicx} % Required for inserting images
\def\ppbar{$\mathrm{p}\overline{\mathrm{p}}$}

\usepackage{booktabs}
\usepackage{algorithm}
\usepackage{algorithmic}
\usepackage{siunitx}
\usepackage{silence}

\usepackage{hyperref}

\begin{document}

\begin{frontmatter}

\journal{{arXiv:2406.19180 }, %The Chinese version is published in 
Chinese Science Bulletin 69 (2024) 4620, \href{https://doi.org/10.1360/TB-2024-0578}{doi:10.1360/TB-2024-0578}}

\title{Protonium: Discovery and Prediction}
\author{Bo-Qiang Ma}
\affiliation{organization={School of Physics, Peking University},
            city={Beijing 100871},
            country={China}}
\date{June 2024}

\begin{abstract}
The Beijing Spectrometer (BESIII) Collaboration reconstructed the invariant mass of three pairs of positive and negative pions by studying the decay process of charmonium to a photon and three pairs of positive and negative pions. They discovered the resonant structures X(1840) and X(1880), 
which are interpreted as the predicted proton-antiproton bound states, also known as protonium. This article briefly introduces the experimental discovery processes of these resonant structures and discusses the theoretical explorations inspired by them. The predictions proposed by these theoretical explorations offer a new perspective for studying the nature of these particles and new decay modes. 
Therefore the collaborative exploration of experiments and theory plays a positive role in deepening understanding of the fundamental laws of nature.       
\end{abstract}


\end{frontmatter}

%\maketitle

\onecolumn

On April 9, 2024, the Beijing Spectrometer III (BESIII) collaboration published a paper in the 
%international physics 
journal ``Physical Review Letters" \cite{BESIII:2023vvr}, reporting the analysis of 100 billion charmonium events produced by the Beijing Electron Positron Collider (BEPC). The study focused on the decay process of charmonium into a photon and three pairs of positive and negative pions. By reconstructing the invariant mass of the three pairs of positive and negative pions and fitting the data with two resonant structures X(1840) and X(1880), 
%a deviation 
with a statistical significance exceeding 10 standard deviations was found for X(1880) with mass and width of 
~$M=1882.1\pm 1.7\pm 0.7$~MeV/$c^2$ and $\Gamma=30.7 \pm 5.5\pm 2.4$~MeV/$c^2$. The mass and width of X(1840) were determined to be 
~$M= 1832.5 \pm 3.1 \pm 2.5$~MeV/$c^2$ and $\Gamma=80.7 \pm 5.2 \pm 7.7$~MeV/$c^2$, consistent with the results obtained by the BESIII collaboration in 2013 based on 200 million charmonium events \cite{BESIII:2013sbm}. This work confirms the existence of X(1840) and X(1880) as new particles in experimental physics. These particles can be interpreted as predicted proton-antiproton bound states \cite{Yan:2004xs}, also known as protonium.

\begin{figure}[htb]
  \centering
  \includegraphics[width=0.7\linewidth]{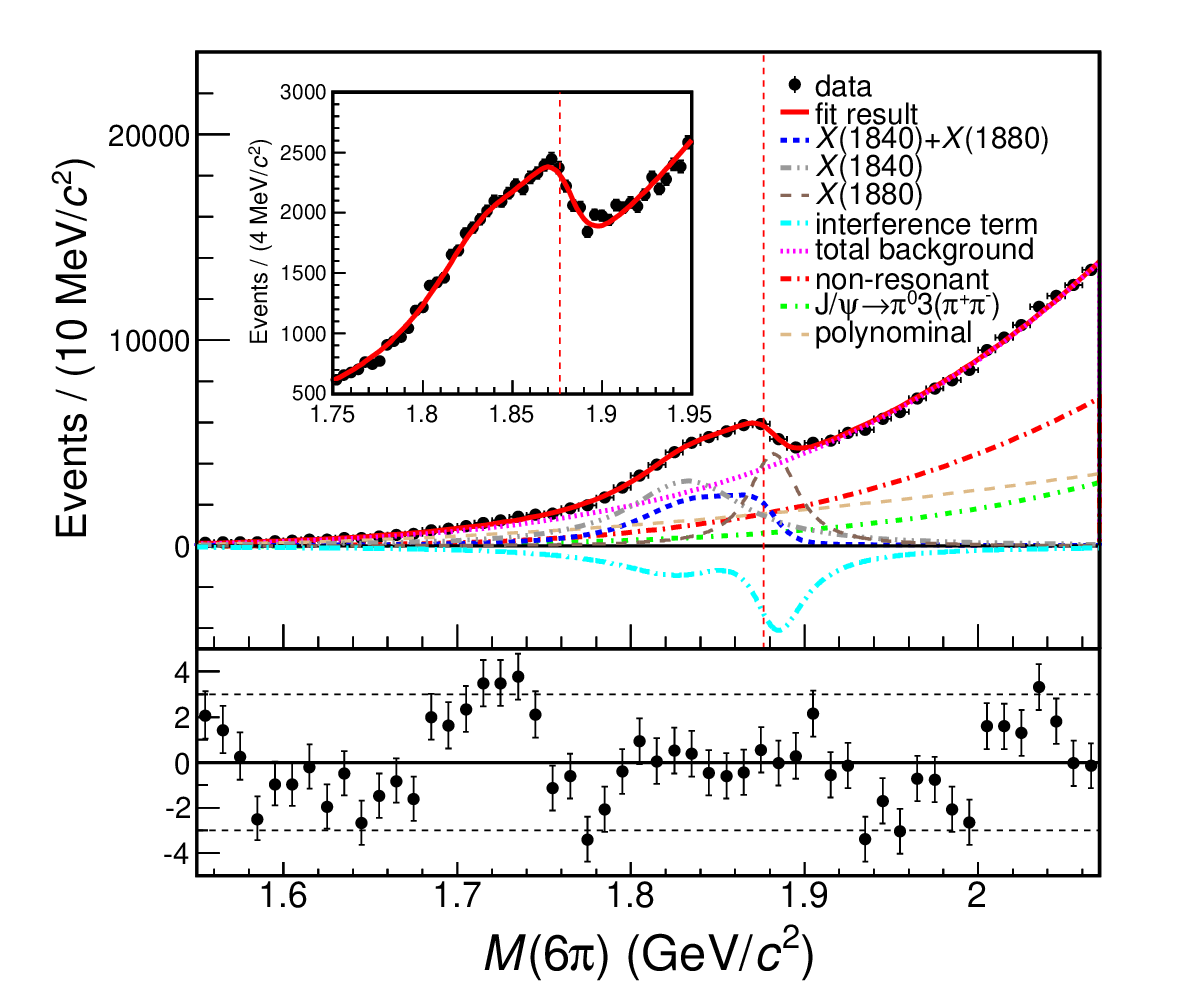}
  %\caption{三对正负介子对的不变质量谱及X(1840)和X(1880)的拟合曲线，选自文献\cite{BESIII:2023vvr}图3(b)。}
  \caption{The invariant mass spectrum of three pairs of positive and negative pions and the fitting curves for X(1840) and X(1880), taken from Figure 3(b) in reference \cite{BESIII:2023vvr}.}
\end{figure}

This BES experiment originated from a research result published by the BES collaboration in ``Physical Review Letters" in 2003 \cite{BES:2003aic}. The study was based on 58 million charmonium events and analyzed the invariant mass spectrum of positive and negative protons near twice the proton mass in the process of charmonium decaying into a photon and a pair of positive and negative protons, revealing a significant near-threshold enhancement phenomenon. By using S-wave analysis to fit the data, a resonant structure with a mass slightly below twice the proton mass was identified, with a mass of 
~$M=1859^{+3}_{-10}~(\mathrm{stat})^{+ 5}_{-25}~(\mathrm{syst})$~MeV/$c^2$ and a width less than 30~MeV/$c^2$.

This result attracted the attention of theoretical physicists, who proposed various explanations to understand the experimental findings. These explanations include interpreting the resonant structure as a bound state X(\ppbar) composed of a proton and antiproton \cite{Yan:2004xs, Datta:2003iy}. Although the central mass of this resonance is lower than the sum of the proton and antiproton masses, the resonance has a width of 30~MeV/$c^2$, allowing for a certain phase space to decay into a pair of free proton-antiproton states, leading to the experimentally observed near-threshold enhancement phenomenon.
In addition, there are also studies to have interpreted X(1859) as a glueball state \cite{Rosner:2003bm, HAO200653}, or as a manifestation of final state interactions \cite{Zou:2003zn}, or as coherent contributions from different scattering channels \cite{Liu:2009vm}, or as phenomena resulting from quark fragmentation \cite{Rosner:2003bm}, and so on.

\begin{figure}[htb]
  \centering
  \includegraphics[width=0.6\linewidth]{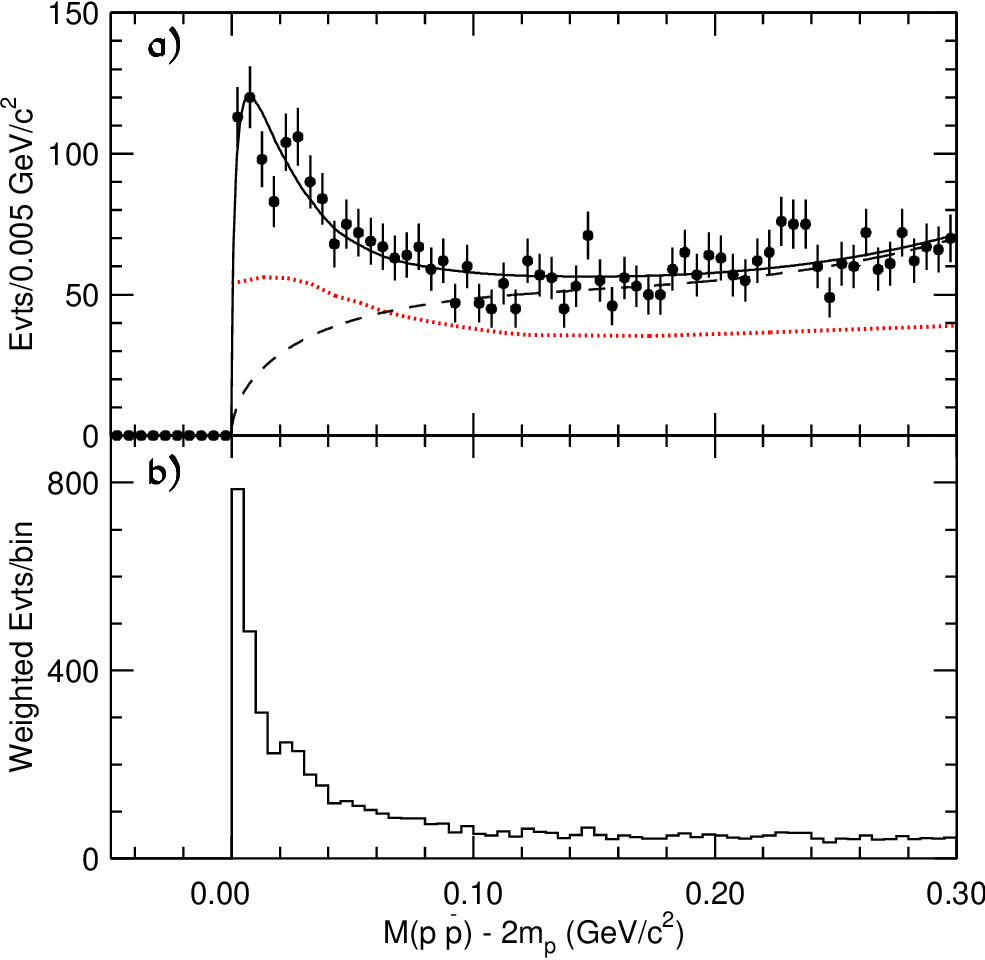}
  %\caption{BES合作组观测到$J/\psi$衰变到光子和质子反质子过程中末态质子反质子质量谱在近域附近的增强现象，选自文献\cite{BES:2003aic}图3。}
\caption{BES Collaboration observed an enhancement phenomenon in the near-threshold region of the final state proton-antiproton mass spectrum in the decay of 
$J/\psi$ to a photon and proton-antiproton process, taken from Figure 3 in reference \cite{BES:2003aic}.}
\end{figure}

In 2005, the BES collaboration observed a resonance structure X(1835) with a mass of 
$M=1833.7\pm 6.1~(\mathrm{stat})\pm 2.7~(\mathrm{syst})$~MeV/$c^2$
  in the decay channel of charmonium 
$J/\psi$ to a photon plus an 
 $\eta'$ meson and a pair of positive and negative 
$\pi$
mesons by reconstructing the invariant masses of the 
 $\eta'$ meson and the pair of positive and negative 
$\pi$ mesons \cite{BES:2005ega}. This structure was further confirmed in subsequent analyses with larger datasets \cite{BESIII:2010gmv}. Based on more data, the BES collaboration also reanalyzed the invariant mass spectrum of X(\ppbar) in the decay channel of charmonium decaying into a photon and a pair of positive and negative protons \cite{BESIII:2010vwa, BESIII:2011aa}. Taking into account final state interactions \cite{Zou:2003zn} and reaction coherence \cite{Liu:2009vm}, they found that the mass of X(\ppbar) also shifted to around 1832~MeV/$c^2$, and there were complex structures near 
$M=1860$~MeV/$c^2$ as well \cite{Jin:2021vct}. These cases demonstrate that the interaction between theory and experiment plays a positive role in the experimental analysis of data.

Professor Mu-Lin Yan from the University of Science and Technology of China and his collaborators made significant contributions to theoretical explorations in explaining the X(\ppbar) structure using proton-antiproton bound states \cite{Yan:2004xs}. British scientist Skyrme attempted to explain the proton and its properties using solitons of chiral fields \cite{Skyrme:1961vq}. Witten argued that the picture of baryons as solitons is consistent with Quantum Chromodynamics (QCD) in the large $N_c$ limit \cite{Witten:1983tw, Witten:1983tx}. In reference \cite{Yan:2004xs}, starting from the theoretical basis of understanding the proton as a Skyrmion, Yan et al. discussed a system composed of Skyrmion and anti-Skyrmion, finding a bound state solution of Skyrmion and anti-Skyrmion system that can explain the bound state of proton and antiproton. They then constructed a phenomenological potential model for proton and antiproton, which could be adopted to calculate the mass and width of the bound state by adjusting parameters, interpreting X(\ppbar) as a bound state of protons and antiprotons, namely a protonium. Additionally, based on experimental phenomena of multi-meson production in low-energy proton-antiproton collisions, Yan et al. explicitly predicted another decay channel of X(\ppbar) as a proton-antiproton bound state, decaying into 4 to 7 mesons instead of 2-3 mesons. The recent analysis by the BESIII collaboration of the decay process of charmonium into a photon and three pairs of positive and negative pions can be seen as a specific realization of the decay of proton-antiproton bound states into 6 mesons. The experimental results are consistent with this theoretical prediction. In turn, this theoretical prediction strengthens the viewpoint of interpreting the resonant states X(1840) and X(1880) discovered by BESIII as protonium.

\begin{figure}[htbp]
  \centering
  {\includegraphics[width=0.45\linewidth]{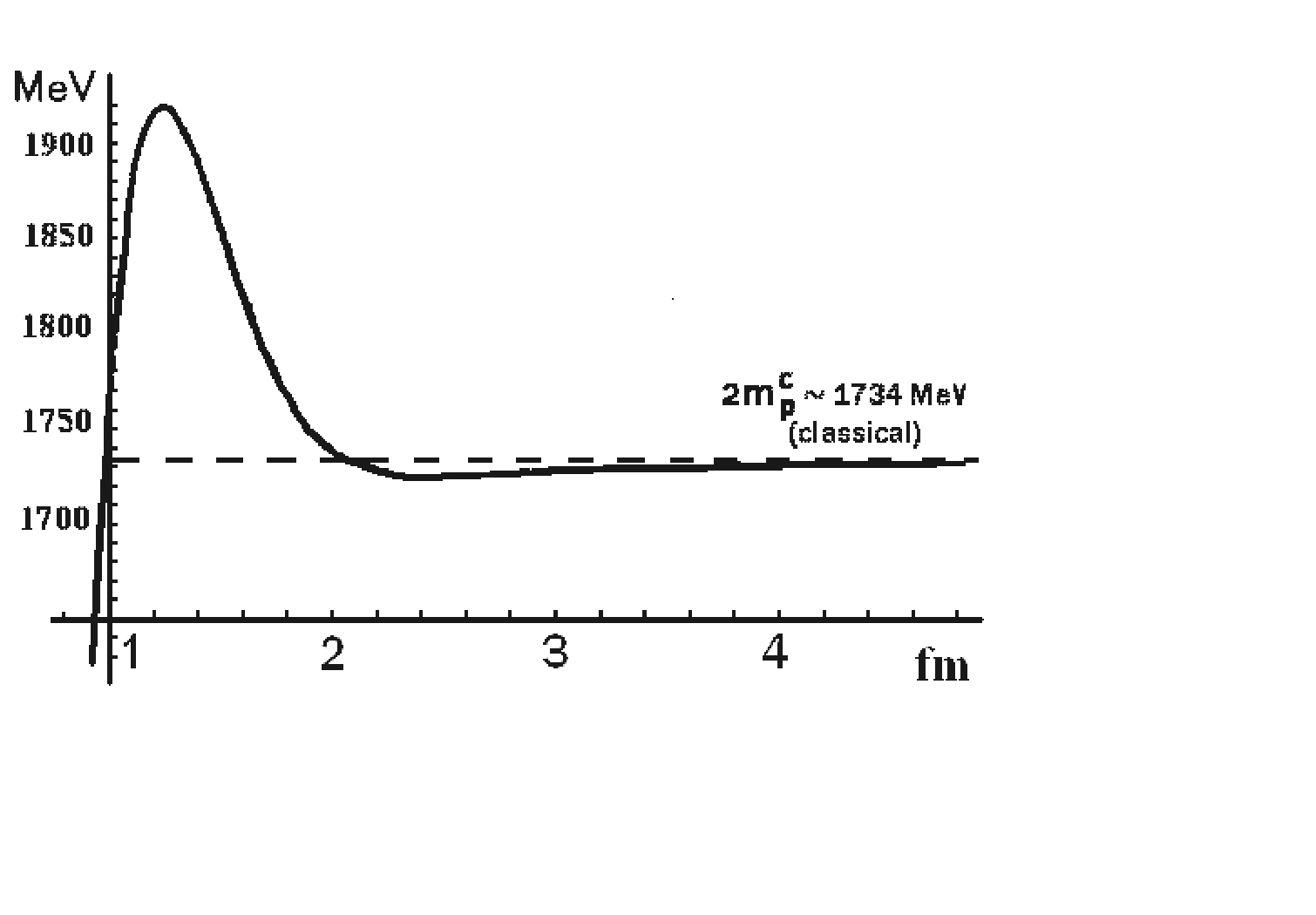}}
  {\includegraphics[width=0.45\linewidth]{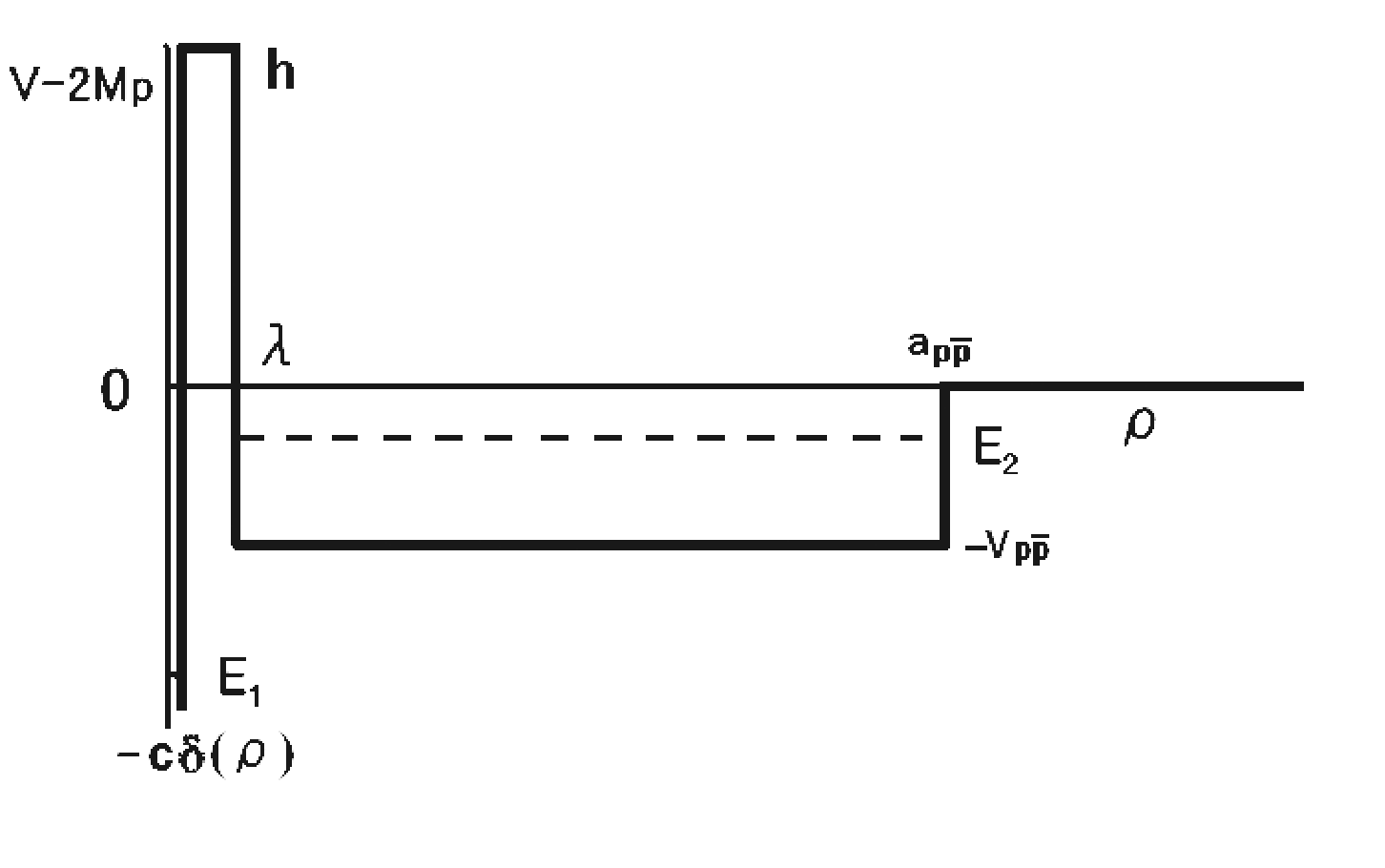}}
  %\caption{左图为斯格明子和反斯格明子组成的系统的静态能量，右图为质子和反质子系统的势模型，分别选自文献\cite{Yan:2004xs}的图1和图2。} 
  \caption{The left figure shows the static energy of a system composed of a skyrmion and anti-skyrmion, while the right figure displays the potential model of a proton and an antiproton system, taken from Figures 1 and 2 in reference \cite{Yan:2004xs}.}
  
\end{figure}

Theoretical work on proton-antiproton bound states can be traced back to 1949 when Fermi and Yang attempted to interpret mesons as deeply bound states of protons and antiprotons \cite{Fermi:1949voc}. Nambu and Jona-Lasinio, based on chiral symmetry, predicted the existence of a protonium with a mass slightly lower than twice the proton mass, in addition to the relatively light pion \cite{Nambu:1961tp, Nambu:1961fr}. The establishment of the quark model in 1964 upended the idea of using meson and nucleon degrees of freedom to understand hadrons. However, theoretical and experimental explorations for finding protonium have never ceased, such as searching for the Coulomb bound state of protons and antiprotons through low-energy proton-antiproton reactions \cite{Klempt:2002ap}. The BES collaboration's discovery of near-threshold enhancement in the invariant mass spectrum of protons and antiprotons in 2003 opened a new era in experimentally discovering hadronic molecular states of protons and antiprotons formed through strong interactions. The theoretical explorations stimulated by experiments provided new insights for the BESIII collaboration to explore protonium, offering stronger evidence for research in this field.

As one of the internationally leading facilities for electron-positron collisions, the Beijing Spectrometer plays a crucial role in experimental research on exotic hadron states or multiquark states. This includes the discovery of various possible tetraquark states, pentaquark states, and even glueball states. X(1840) and X(1880) may belong to a special category of exotic hadron states, specifically protonium composed of a proton and antiproton, or a hexaquark state consisting of 3 quarks and 3 antiquarks. The discovery by the BESIII collaboration may have revealed the existence of protonium for the first time. However, this conclusion still requires further experimental research to measure the properties of X(1840) and X(1880) more accurately and to compare them with theoretical predictions.

 Chinese theoretical physicists have conducted extensive theoretical research on the theoretical foundation of proton-antiproton bound states, as well as extending the concept to other baryons, thus exploring the direction of baryon-antibaryon bound states \cite{Ding:2005ew, Ding:2005gh, Ding:2005tr, Chang:2004us, He:2004ih, Zhu:2005ns, Liu:2004er, Wang:2006sna}. This has provided new opportunities for experimental explorations of protonium and even baryonium. With more accumulation of experimental data and increasingly precise theoretical predictions, it is believed that the BESIII collaboration will make more discoveries in the experimental exploration of protonium and baryonium, making greater contributions to a deeper understanding of the fundamental laws of the natural world.

%\begin{acknowledgments}
%  致谢内容。
%\end{acknowledgments}

%\usepackage[backend=biber,style=authoryear]{biblatex}
%\addbibresource{protonium.bib}

% 将文献作者姓名的字母全部大写
%\DeclareNameFormat{author}{%
%  \nameparts{\uppercase{#1}}%
%  \usebibmacro{name:family-given}
%  \usebibmacro{name:andothers}
%}

%或者仅将文献作者姓名的首字母大写
%\DeclareNameFormat{author}{
%  \nameparts{\namepartfamily}
%  \namepartgiven{\mkbibnamegiven{\MakeUppercase{\namepartgiven}}}
%  \namepartprefix{\namepartprefix}
%  \nameparts{\namepartfamily}
%}

%\bibliographystyle{cjc}
%\bibliographystyle{csb2}
\bibliographystyle{csb}
\bibliography{protonium}

%\makebiographies

%作者说明：文章中出现的ppbar应该排版为p\overline{\mathrm{p}}
 
\end{document}